\documentclass[%
 aip, 
 amsmath,amssymb,
 reprint,%
]{revtex4-1}

\usepackage{graphicx}% Include figure files
\usepackage{dcolumn}% Align table columns on decimal point
\usepackage{bm}% bold math
\usepackage[mathlines]{lineno}% Enable numbering of text and display math
\usepackage[normalem]{ulem}
\usepackage{color}

\usepackage[utf8]{inputenc}
\usepackage[T1]{fontenc}
\usepackage{mathptmx}
\usepackage{etoolbox}
\usepackage{color}
\usepackage[mathlines]{lineno}

\makeatletter
\def\@email#1#2{%
 \endgroup
 \patchcmd{\titleblock@produce}
  {\frontmatter@RRAPformat}
  {\frontmatter@RRAPformat{\produce@RRAP{*#1\href{mailto:#2}{#2}}}\frontmatter@RRAPformat}
  {}{}
}%
\makeatother
\begin{document}

\preprint{AIP/123-QED}

\title{Characterization of Low-noise Backshort-Under-Grid Kilopixel Transition Edge Sensor Arrays for PIPER}
% Force line breaks with \\

\author{Rahul~Datta}
\affiliation{%
The William H. Miller III Department of Physics and Astronomy, Johns Hopkins University,
3701 San Martin Drive, Baltimore, MD 21218, USA
}%
%\email{rdatta2@jhu.edu} %% email address is required

\author{Sumit Dahal}
\affiliation{%
NASA Goddard Space Flight Center, 8800 Greenbelt Road, Greenbelt, MD 20771, USA
}%
\author{Eric~R.~Switzer}
\affiliation{%
NASA Goddard Space Flight Center, 8800 Greenbelt Road, Greenbelt, MD 20771, USA
}%
\author{Regis~P. Brekosky}
\affiliation{%
NASA Goddard Space Flight Center, 8800 Greenbelt Road, Greenbelt, MD 20771, USA
}%
\author{Thomas~Essinger-Hileman}
\affiliation{%
NASA Goddard Space Flight Center, 8800 Greenbelt Road, Greenbelt, MD 20771, USA
}%
\author{Dale~J.~Fixsen}
\affiliation{%
NASA Goddard Space Flight Center, 8800 Greenbelt Road, Greenbelt, MD 20771, USA
}%
\author{Christine~A.~Jhabvala}
\affiliation{%
NASA Goddard Space Flight Center, 8800 Greenbelt Road, Greenbelt, MD 20771, USA
}%
\author{Alan~J.~Kogut}
\affiliation{%
NASA Goddard Space Flight Center, 8800 Greenbelt Road, Greenbelt, MD 20771, USA
}%
\author{Timothy~M.~Miller}
\affiliation{%
NASA Goddard Space Flight Center, 8800 Greenbelt Road, Greenbelt, MD 20771, USA
}% 
\author{Paul~Mirel}
\affiliation{%
NASA Goddard Space Flight Center, 8800 Greenbelt Road, Greenbelt, MD 20771, USA
}%
\author{Edward~J.~Wollack}
\affiliation{%
NASA Goddard Space Flight Center, 8800 Greenbelt Road, Greenbelt, MD 20771, USA
}%

%\homepage[]{Your web page}
%\thanks{}
%\altaffiliation{}
%\affiliation{affiliation}

\date{\today}% It is always \today, today,
             %  but any date may be explicitly specified

\begin{abstract}
We present laboratory characterization of kilo-pixel, filled backshort-under-grid (BUG) transition-edge sensor (TES) arrays developed for the Primordial Inflation Polarization ExploreR (PIPER) balloon-borne instrument. PIPER is designed to map the polarization of the CMB on the largest angular scales and characterize dust foregrounds by observing a large fraction of the sky in four frequency bands in the range 200–600 GHz. The BUG TES arrays are read out by planar SQUID-based time division multiplexer chips (2dMUX) of matching form factor and hybridized directly with the detector arrays through indium bump bonding. Here, we discuss the performance of the 2dMUX and present measurements of the TES transition temperature, thermal conductance, saturation power, and preliminary noise performance. The detectors achieve saturation power $< 1$\,pW and phonon noise equivalent power (NEP) on the order of a few aW/$\sqrt{\rm Hz}$. Detector performance is further verified through pre-flight tests in the integrated PIPER receiver, performed in an environment simulating balloon float conditions. 

\end{abstract}


%\pacs{65.40.De}% insert suggested PACS numbers in braces on next line

\maketitle %\maketitle must follow title, authors, abstract and \pacs

% Body of paper goes here. Use proper sectioning commands.
% References should be done using the \cite and \label commands.

\section{Introduction}

Bolometric detectors employing Transition-edge sensors (TES) measure power deposited by a flux of photons using the temperature dependence of a superconductor’s resistance when biased in transition. Large arrays of TES bolometers have been widely deployed in astronomical instruments operating from the ground~\cite{Niemack2008, 10.1117/12.2232986,2018SPIE10708E..1ZD,2016JLTP..184..805S,dahal2020} and suborbital platforms~\cite{2015apra.prop...17R, 2018JAI.....740008H, Bergman2018, 2018ApJS..239....8E} requiring background-limited sensitivity across millimeter and submillimeter wavelengths. Ultra low-noise (on the order of $10^{-19}$ W/$\sqrt{\rm Hz}$), high sensitivity TES bolometer arrays were developed to support the SAFARI far-infrared instrument concept~\cite{2009AIPC.1185...42K, 10.1117/12.2233472, Ridder2016, Suzuki2016}. TES bolometers are a promising candidate for planned next-generation space satellites such as LiteBIRD~\cite{2022SPIE12190E..0IW,2020SPIE11443E..2GM}, and the proposed PICO and Origins Space Telescope, but require further development to meet the exquisite stability and sensitivity requirements across near-, mid-~\cite{2018JAI.....740015R}, and far-infrared~\cite{10.1117/1.JATIS.7.1.011005, 2017arXiv170902389F}.

The TES bolometer is typically operated at near-constant voltage bias to maintain the superconductor on a stable transition via negative electrothermal feedback. This superconductor on an isolated island, biased on transition and coupled to a thermal mass with a heat capacity, measures changes in temperature of the thermal mass indirectly by measuring how much electrothermal feedback power is required to maintain a near-constant temperature. The isolated island is connected to a bath at a lower temperature through a thermal link with a conductance to define the bolometer time constant. For a given bath temperature and $T_{c}$, the thermal conductance to the bath sets the device saturation power and noise. Isolation via long (diffusive) legs is one technique used to achieve low noise through control of fluctuations driven by phonons. Bolometers employing long legs provide a natural path to realize the device parameters for absorber coupled sensors considered here. In the alternate approach employing transmission line coupled sensors \cite{rostem2016}, ballistic, phononic, and other structures may be more suitable for achieving low noise from a mechanical and thermal transport perspective~\cite{doi:10.1063/1.4869737}.

\begin{figure}
\begin{center}
\includegraphics[width=0.85\linewidth,keepaspectratio]{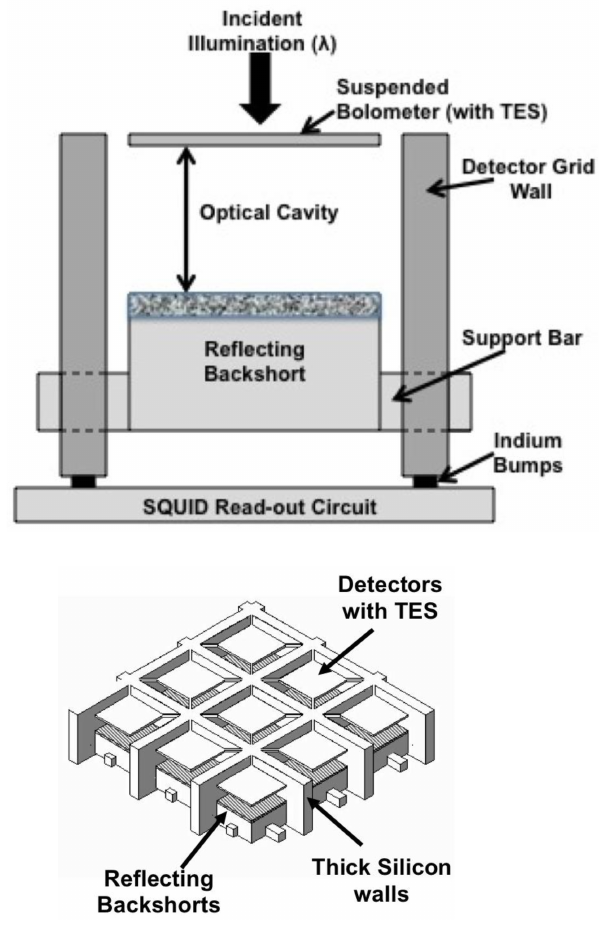}
\end{center}
\caption{{\it{Top:}} Schematic representation of the BUG detector architecture showing the cross-section of a single pixel in the BUG array. {\it{Bottom:}} An array of leg-isolated bolometers with reflective backshorts inserted behind the pixels. Figure adapted from [\!\citenum{10.1117/12.2056995}].}
\label{fig:bug_arch}
\end{figure}

The scalable backshort-under-grid~\cite{10.1117/12.2056995, 10.1117/12.856491} (BUG) detector architecture supports large arrays of low-noise TES bolometers. The BUG assembly consists of three individual components -- a TES bolometer array, a quarter-wavelength backshort grid, and a multiplexing readout merged into a single working unit. By adjusting the spacing between the detector grid and the reflective backshort~\cite{Carli:81}, these BUG arrays can be optimized for a wide range of infrared to millimeter applications requiring large format background-limited detectors~\cite{Allen2006, Sharp2008, 2015apra.prop...17R, Harper2018}. In this paper, we describe the laboratory characterization of BUG detector arrays with low noise and low saturation power designed for high sensitivity observations of the CMB sky from a balloon platform. These arrays developed for the Primordial Inflation Polarization ExploreR (PIPER)~\cite{10.1117/12.857119,2014SPIE.9153E..1LL,10.1117/12.2231109,10.1117/12.2313874} comprise a filled, $32\times40$ grid of suspended TES bolometers. For the PIPER BUG arrays, the backshort behaves more like a choke than an infinite reflective plane with loading from dielectric walls of the detector frame (see Fig.~\ref{fig:bug_arch}), this changes the delay and can be used to increase the absorption bandwidth. The backshort spacing was optimized through numerical simulations incorporating the details of the back-termination and the dielectric walls, which influence the device absorption bandwidth. This approach allows single focal plane arrays to be used for all the PIPER frequency bands.

PIPER is a balloon-borne instrument designed to map the polarization of the CMB on the largest angular scales and characterize dust foregrounds by observing a large fraction of the sky in four frequency bands centered at 200, 270, 350, and 600 GHz. PIPER uses twin telescopes with variable-delay polarization modulators~\cite{2014RScI...85f4501C} (VPMs). These front-end VPMs provide polarization sensitivity and systematic control by modulating linear to circular polarization at 3--5\,Hz. Twin co-pointed telescopes with VPM modulation and an analyzer grid enable instantaneous measurement of Stokes $Q$ and $U$ for each pointing. The instrument design reaches background-limited sensitivity provided by fully cryogenic ($<5$\,K) optics that focus the sky signal onto kilo-pixel BUG TES arrays at the focal plane of each telescope, which are cooled to $100$\,mK using a continuous adiabatic demagnetization refrigerator~\cite{10.1117/1.JATIS.4.2.021403,2019RScI...90i5104S} (CADR). 

\begin{figure*}[ht]
\begin{center}
\includegraphics[width=0.85\linewidth,keepaspectratio]{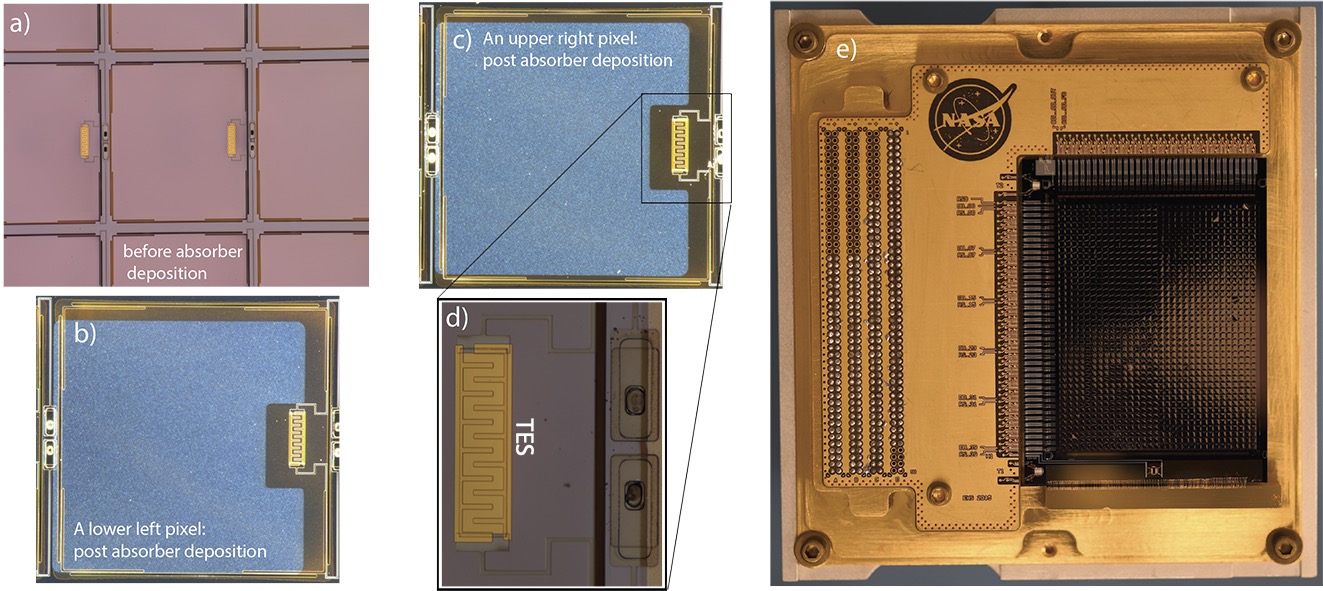}
\end{center}
\caption{A $32\times40$ BUG TES array designed for PIPER. {\it{(a)}}: Close-up image showing part of a fabricated detector wafer before the Bi absorber was deposited; {\it{(b)}}: Image of a single pixel from the lower left part of the wafer after absorber (indicated by blue) deposition; {\it{(c)}}: Image of a single pixel from the upper right part of the wafer after absorber deposition. For this array, the absorber mask was slightly misaligned inadvertently, which resulted in the absorber overlapping with the legs of the TES as seen in both images; {\it{(d)}}: Close-up image of a TES; {\it{(e)}}: Photo of a packaged detector wafer.}
\label{fig:arr_imag}
\end{figure*}

This paper is organized as follows. Section~\ref{sec:detarch} presents an overview of the detector and readout architecture. Section~\ref{sec:2dMUX_screening} reviews measurements of the 2dMUX chip and tuning of multiplexing parameters. Measurements of the thermal parameters of the BUG TES arrays are presented in Section~\ref{sec:det_array_params}. We also discuss pre-flight characterization and verification of detector performance from the PIPER integrated receiver test performed in a liquid helium (LHe) test dewar simulating float conditions in Section~\ref{sec:det_array_params}. Finally, we summarize in Section~\ref{sec:conclusion}.

%\section{Requirements}

%…this is not the exact electromagnetic design used in PIPER, but it is a useful reference which discussed a reflective termination with a quarter wave design (notice for PIPER the backshort is actually more of a choke than an infinite reflective plane with dielectric loading from the wall – this changes the delay and can be used to increase the absorption bandwidth). This said, at a high level – this would be a useful starting point to explain how the absorber membranes works and link it to prior art found in the literature…

%If one desired to motivate this approach the paper by Carli and lorio-Fili (1981)~\cite{Carli:81} could be cited, however, in practice the coupling configuration was optimized through numerical simulation (i.e., to include the details of the back-termination and the dielectric walls of the detector frame – these factors influences the choice of the back-termination spacing and the device absorption bandwidth in a non-trivial manner – the termination structure is not a simple resonant circuit set by a single quarter wave delay).

\section{Overview of Detectors and Readout}
\label{sec:detarch}

Each pixel of a PIPER BUG array, comprising a planar absorbing element thermally coupled to a TES, measures total incident power over a frequency band defined by bandpass filters in front of the array. Fig.~\ref{fig:bug_arch} shows a schematic of the BUG architecture.  Each BUG detector assembly consists of 32$\times$40 square pixels with a pixel-to-pixel spacing of 1123.5 $\mu$m and a fill factor of $\sim$95\%. The TESs are on leg-isolated, absorber-coated, suspended silicon membranes and designed to have a superconducting transition ($T_{\rm c}$) at 140~mK. The electrical signals for the TES bolometers are routed to the back of the grid structure, where indium bump bonds implement a superconducting connection to wrap-around vias~\cite{2016JLTP..184..615J}. A second grid consisting of an array of reflecting elements is nested into the back of the first grid to provide a backshort for each detector as shown in the bottom panel of Fig.~\ref{fig:bug_arch}. In the case of PIPER, the backshort distance was chosen to maximize detector absorption over the lowest frequencies that contain the CMB signal. The BUG array is hybridized via indium-bump-bonding with two-dimensional superconducting quantum interference device (SQUID) time-division multiplexer (TDM) chip~\cite{doi:10.1063/1.1593809} (2dMUX) developed by NIST for SCUBA-2~\cite{2013MNRAS.430.2513H}, achieving a multiplexing factor of 40. The device absorber membrane is coated with a thin absorbing film of bismuth, with a surface resistance of 377\,ohms/square for impedance-matched coupling to free space as shown in Fig.~\ref{fig:arr_imag}.

The TDM architecture uses two stages of SQUID amplification~\cite{doi:10.1063/1.123255,doi:10.1063/1.1593809}. Each TES in the array is inductively coupled to its own first-stage DC SQUID series array (SQ1). The series array is connected in parallel with a Josephson junction array in the form of a Zappe interferometer~\cite{1059358} which acts as a flux-activated row select switch (RS). Each SQ1/RS pair forms a channel of the readout, with many channels connected in series to form columns of channels. A bias current is applied to the chain of SQ1/RS units in parallel with a 1 ohm resistor. The dynamic resistance of the SQ1/RS chain is large compared to 1 ohm, so it is approximately voltage-biased. All but one RS are left in the superconducting state, so the current does not pass through their companion SQ1s. One RS associated with the addressed channel is flux-biased in its normal state, so the total voltage is dropped across the selected SQ1/RS unit. The 2dMUX has 32 columns and 41 rows, the extra row being a ``dark'' row that is not connected to any TES so that it can be used to characterize readout noise and magnetic field pickup. During operation, the detectors are set in transition by applying an appropriate bias voltage across a detector bias line. There are a total of 20 detector bias lines with each line being used to bias 2 rows of detectors. This allows selecting independent bias voltages optimized for each group of 80 detectors.

%{\textcolor{red}{(would have been good to show measured NEP$_{phonon}$, but not sure that this expected NEP$_{phonon}$ plot is adding any value)}}
The warm readout utilizes one Multi-Channel Electronics (MCE)~\cite{2008JLTP..151..908B} for each array. Each MCE crate contains nine cards; one addressing card (AC), three bias cards (BCs), four readout cards (RCs), and one clock card (CC). The AC contains 41 digital-to-analog converters (DACs) for switching rows in the 2dMUX, each BC contains 32 DACs for setting and switching SQUID biases and feedback, the RCs implement the flux-locked loop (FLL) used to maintain SQUID linearity and read out the TESs, and the CC serves as an interface for each MCE. Data synchronization is accomplished using a sync box which provides timing signals to all MCEs and the telescope housekeeping system, allowing detector and pointing information to be synchronously read out.

\section{2D Multiplexer chip screening}
\label{sec:2dMUX_screening}
\begin{figure}[h]
\begin{center}
\includegraphics[width=1\linewidth,keepaspectratio]{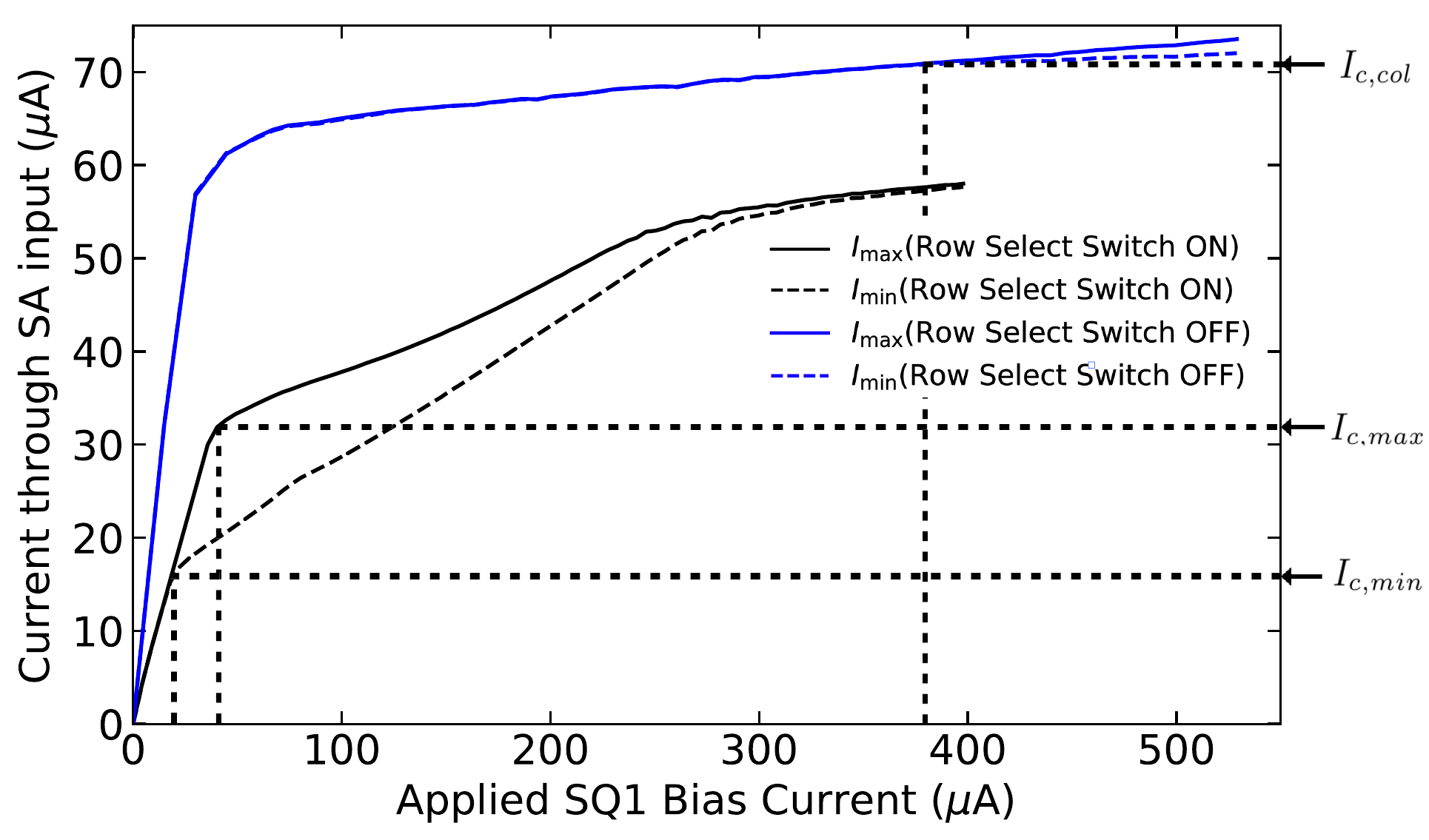}
\end{center}
\caption{The minimum (dashed line) and maximum (solid line) current through an example multiplexer channel's SQUID (SQ1) and row select switch (RS), maximized over all values of applied SQ1 feedback, measured from SQ1 servo curves as a function of SQ1 bias. Measurements are shown both with the RS superconducting (blue) and normal (black). The vertical dashed lines represent various applied SQ1 bias currents resulting in corresponding critical currents as indicated by the horizontal dashed lines. The critical currents $I_{\rm c,min}$ and $I_{\rm c,max}$ are defined here when the RS is normal. $I_{\rm c,min}$ is the smallest current applied to the SQ1 bias for which modulation appears, or, where the minimum and maximum current first diverge. $I_{\rm c,max}$ is the current for which the superconducting branch of the SQ1 V-$\phi$ curve first completely disappears, or, where the difference is maximized. Also noted is the SQ1 bias at which the readout column containing this channel first exhibits persistence at current $I_{\rm c,col}$. }
\label{fig:MUX_plot}
\end{figure}

\begin{figure*}
\begin{center}
\includegraphics[width=0.95\linewidth,keepaspectratio]{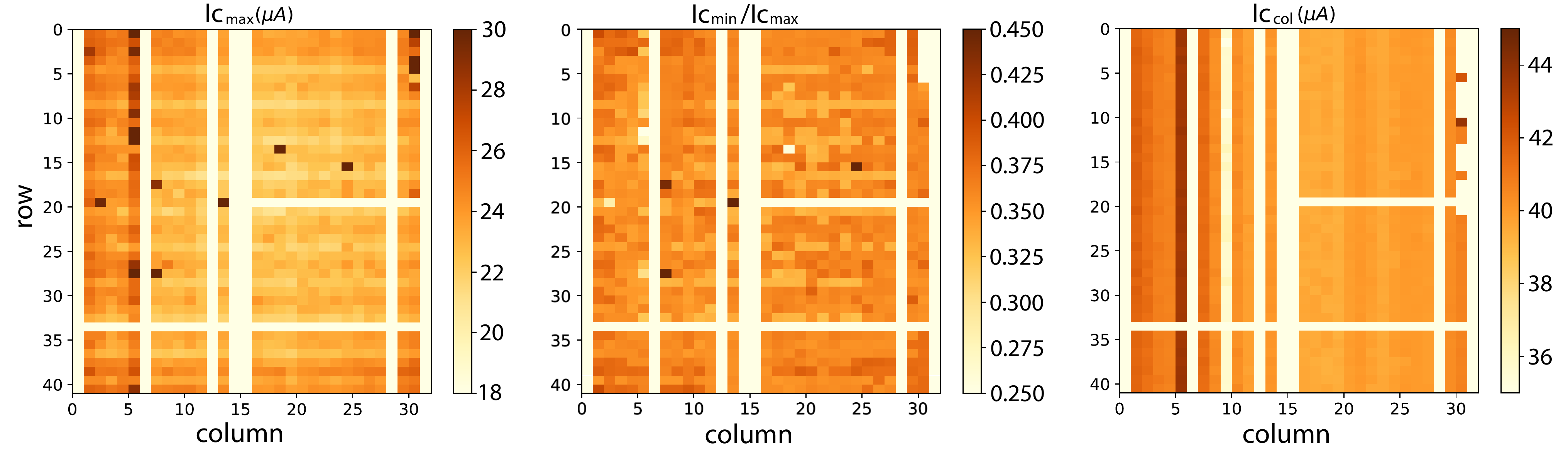}
\end{center}
\caption{Maps of critical currents, $I_{\rm c,max}$ (left), ratio $I_{\rm c,min}$/$I_{\rm c,max}$ (middle), and ratio $I_{\rm c,col}$/$I_{\rm c,max}$ (right) for an accepted 2dMUX that was hybridized with Array~2. Two of the six entirely dead columns were due to bad SA chips, two others were related to issues in the readout chain of the test setup, and the remaining two could potentially be a non-functional 2dMUX column. The completely dead row was also related to an issue in the test setup readout, while the half-dead row was a partially functioning row on the 2dMUX.}
\label{fig:MUX_2dMaps}
\end{figure*}

The 2dMUXs were first screened in an independent test package. We found that use of non-magnetic hardware and printed circuit boards (e.g., without an Ni plating adhesion layer) in the test package was important for obtaining reliable test results and uniform 2D SQUID multiplexer operability. The chips were tuned following the procedure described in [\!\!\citenum{2016SPIE.9914E..1GH}], optimizing the parameters required at each step. Fig.~\ref{fig:MUX_plot} shows a diagnostic plot for a good 2dMUX channel to estimate the various critical currents and optimal SQ1 bias point. Acceptance of the 2dMUX chips requires achieving a 85$\%$ yield satisfying the criteria $I_{\rm c,max} > 2 I_{\rm c,min}$ and $I_{\rm c,col}>2I_{\rm c,max}$, and with curves for the minimum and maximum current through the SA input as a function of applied SQ1 bias current that approximately resembles the shape in Fig.~\ref{fig:MUX_plot}. Fig.~\ref{fig:MUX_2dMaps} shows 2D maps of the various critical currents for an accepted 2dMUX. Most of the dead rows and columns are explained by failures in the readout wiring of the test setup that are unrelated to the 2dMUX. These can be repaired prior to flight.

\section{Characterization of the Kilo-pixel Detector Arrays}
\label{sec:det_array_params}

The hybridized detector arrays were first characterized through cooldowns in a laboratory test cryostat with the detectors in a near light-tight environment. These ``dark tests'' characterize the thermal and electrical properties of the detectors. Dark tests of Array~1 were performed both before and after absorber deposition. After dark characterization the performance of the detectors was verified and further characterized after integration in the PIPER receiver. The integrated receiver test environment \cite{2019RScI...90i5104S} provides both performance and functional verification before flight. Specific tests for the detectors include 1) interaction with the CADR\cite{2019RScI...90i5104S}, 2) verification of thermal stray light control within the receiver, 3) approximation of flight optical loading established through an AN-72 multi-layer carbon-loaded foam Eccosorb~\footnote{https://www.laird.com/products/microwave-absorbers/microwave-absorbing-foams/eccosorb-an/78082099} sheet cooled by a superfluid pump~\cite{Kogut2021}, and placed over each receiver window. Below, we describe the test setups and summarize the results. 

\begin{figure*}
\begin{center}
\includegraphics[width=1\linewidth,keepaspectratio]{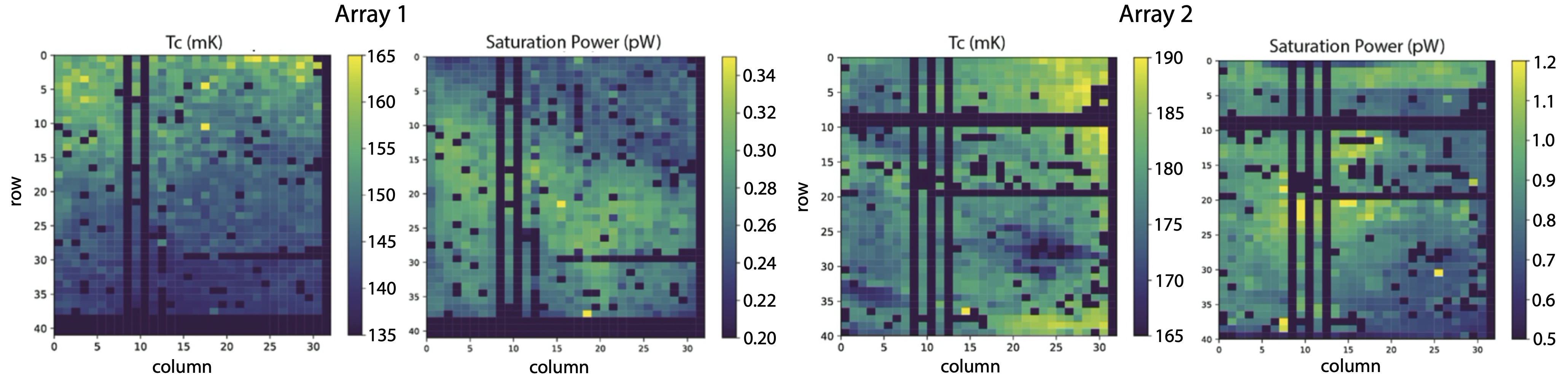}
\end{center}
\caption{Maps of the measured transition temperature (T$_{\rm c}$) and saturation power (P$_{\rm sat}$) of the two arrays (left: Array~1 and right: Array~2) inferred from dark detector tests. The dark lines and squares scattered across the array are dead rows and columns and dead pixels. These are a combination of non-functional 2dMUX channels, dead detectors, and readout wiring failures. The detector yield is 78\% and 72\% for Array~1 and 2, respectively. However, upto 90\% yield is potentially achievable on each array after repairing readout wiring failures. Table~\ref{tab:yield} breaks down the yield for each array.}
\label{fig:Dark_params_maps}
\end{figure*} 

\begin{figure*}
\begin{center}
\includegraphics[width=1\linewidth,keepaspectratio]{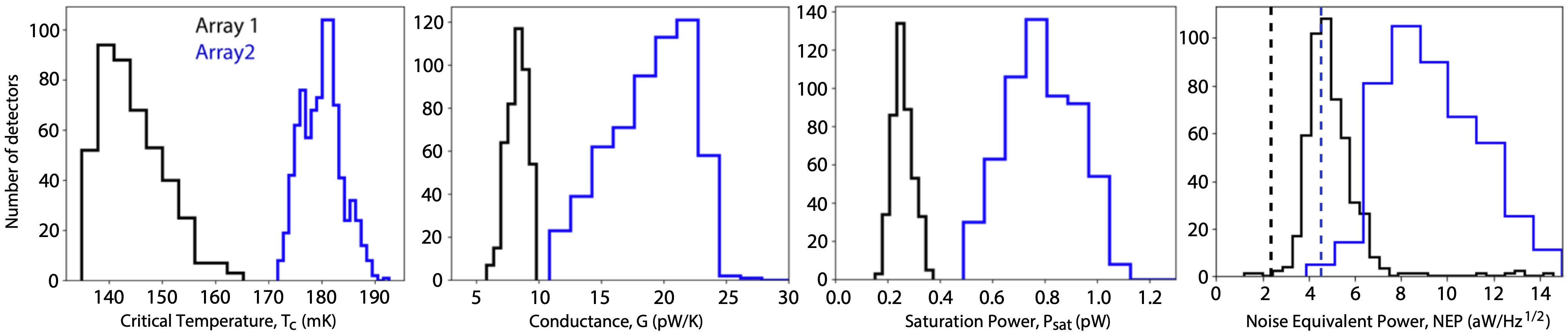}
\end{center}
\caption{Histograms of the measured detector parameters: transition temperature ($T_{\rm c}$), thermal conductance ($G$), saturation power ($P_{\rm sat}$) inferred from dark detector tests in the lab, and noise equivalent power (NEP) inferred from pre-flight integrated receiver tests (see Section~\ref{subsec:int_test}) using Eq.~\ref{eq:nep} for the two arrays (black: Array~1 and blue: Array~2). The array mean phonon NEPs (NEP$\rm _{G}$) are indicated by the vertical dashed lines.}
\label{fig:Dark_params_histograms}
\end{figure*}

\subsection{Dark Detector Tests}

\begin{figure*}
\begin{center}
\includegraphics[width=1\linewidth,keepaspectratio]{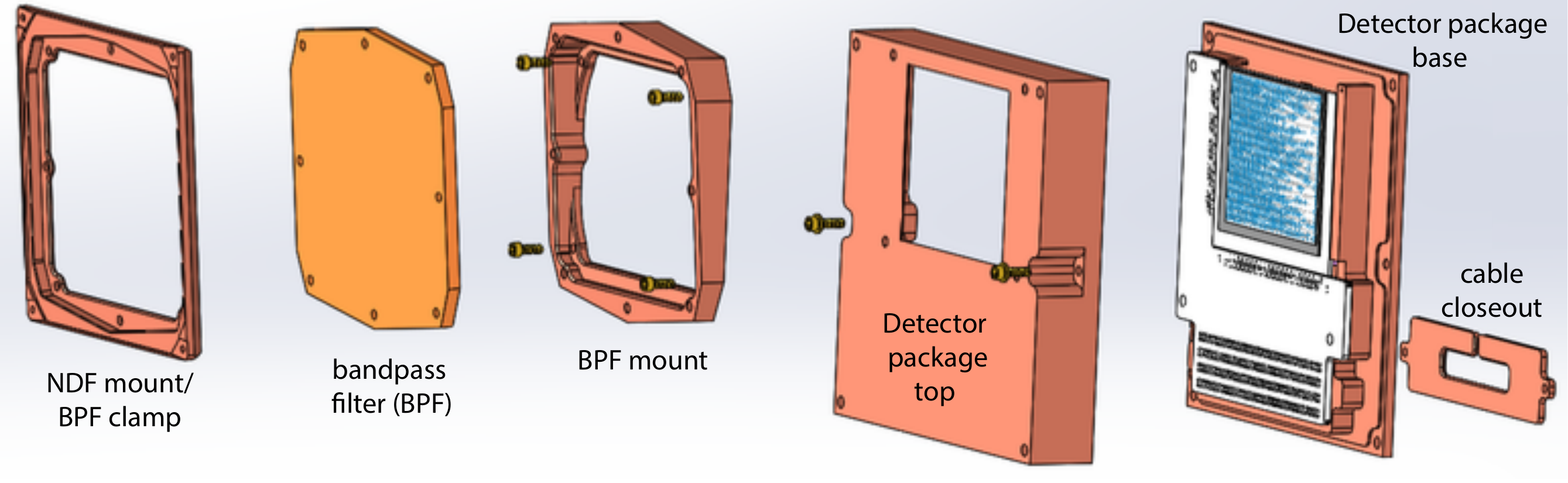}
\end{center}
\caption{Exploded schematic of the detector package. From left to right: the NDF mount also serving as the bandpass filter (BPF) clamp is used to hold down and heat sink the NDF via titanium spring clips at the four corners. The BPF clamp positions the BPF above the detector array and tilted at an angle to limit ghosting. The interiors of the BPF mount and the detector package top part are blackened following the recipe described in~[\!~\citenum{2012SPIE.8452E..3IS}]. The hybridized detector and 2dMUX assembly is mounted on the detector package base along with a readout circuit board. The cable closeout provides an opening for the readout wires to exit the package from behind.}
\label{fig:det_pkg}
\end{figure*}

The detector wafer hybridized with the 2dMUX chip was packaged in a gold-plated copper enclosure, whose interior surfaces were blackened following the recipe described in~[\!~\citenum{2012SPIE.8452E..3IS}]. Fig.\,\ref{fig:det_pkg} shows an exploded view of the detector package. The opening in the package for the band-defining filter was capped off with a copper plate, and the opening for the wires in the cable closeout was carefully taped around the wires, providing a dark environment for the measurement of the detector thermal parameters. The package was cooled to operating temperatures by an adiabatic demagnetization refrigerator (ADR). We define the saturation power $P_{\rm sat}$ as the absorbed optical power required to reach 90\% $R_{\rm N}$, where $R_{\rm N}$ is the TES normal resistance. Joule power applied via detector bias acts as a proxy for absorbed optical power. We measure the saturation power through current-voltage (IV) curves, which first bias the detectors normal and then step down through the superconducting transition. We evaluate thermal conductance performance of the detector legs by a measurement of $P_{\rm sat}$ as a function of bath temperature $T_{\rm b}$. 

The usual power-law dependence~\cite{Irwin2005,Mauskopf_2018} for the power conducted to the thermal bath at temperature $T_{\rm b}$ from the TES at temperature $T$ follows
\begin{equation} 
P(T) = \kappa(T^{n}-T_{\rm b}^{n}),
\label{eq:Psat}
\end{equation}
where $\kappa$ and $n$ are constants. Following Equation~\ref{eq:Psat}, $P_{\rm sat} = P(T=T_{\rm c})$, where $T_{\rm c}$ is the transition temperature of the TES. The thermal conductance to the bath, $G$, is then 
\begin{equation} 
G = \frac{dP}{dT} \biggl |_{T=T_{\rm c}} = \frac{dP_{\rm sat}}{dT_{\rm c}} = n \kappa T_{\rm c}^{n-1}.
\label{eu:cond}
\end{equation}

\begin{figure*}
\begin{center}
\includegraphics[width=1\linewidth,keepaspectratio]{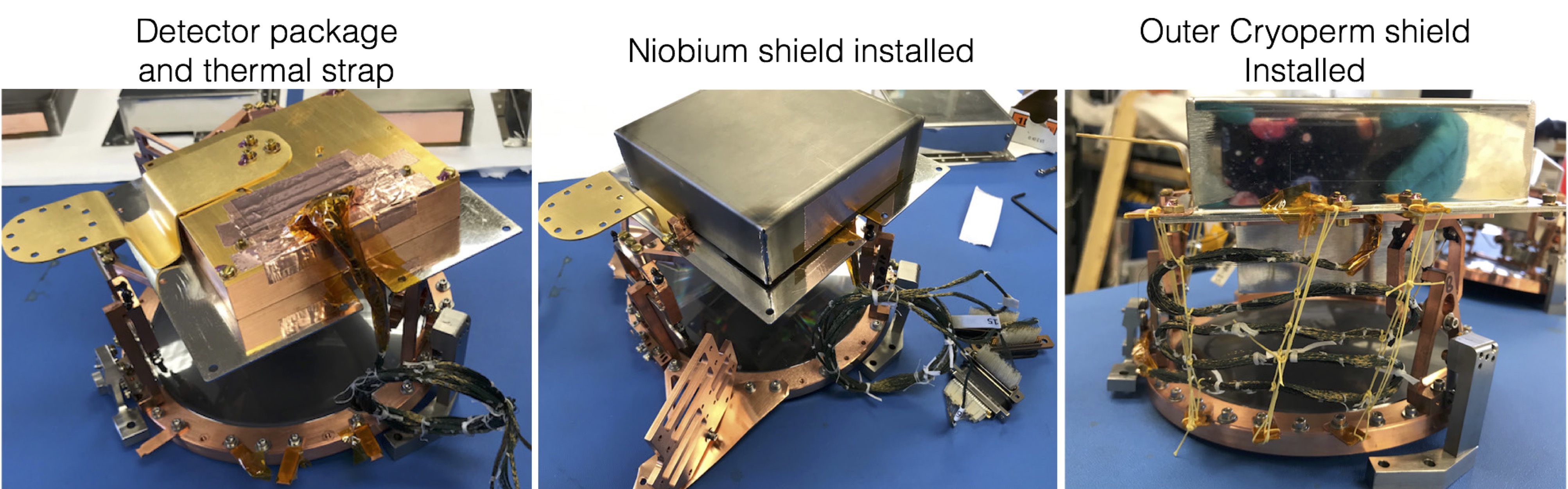}
\end{center}
\caption{Receiver Integration: Each detector package is cooled through a thermal strap that connects to the main 100\,mK CADR bus. The package assembly was capped from behind with an Nb box and enclosed in an Amuneal A4K shroud. The wiring harness was routed by meandering across a Kevlar suspension to reduce the heat leak to the CADR. One of the two silicon lenses that constitute the reimaging optics can be seen in this assembly. Arrays 1 and 2 are both integrated into near-mirror image assemblies and installed inside the receiver cryostat.}
\label{fig:det_int}
\end{figure*} 

Depending on the bath temperature, the bias ramp for the IV curve acquisition, which drives the detectors normal over the whole array, can lead to momentary heating of the bath. A PID control of the ADR cold stage temperature was set up to maintain the bath temperature approximately constant during the course of an IV acquisition. With the measured $P_{\rm sat}$ values at different values of $T_{\rm b}$ and using Eq.~\ref{eq:Psat}, we perform a three-parameter fit to infer $T_{\rm c}$, $\kappa$, and $n$ for each detector. Fig.~\ref{fig:Dark_params_maps} shows array maps of $T_{\rm c}$ inferred from the fits, and measured $P_{\rm sat}$ at the nominal 100~mK operational bath temperature. Fig.~\ref{fig:Dark_params_histograms} shows histograms of $T_{\rm c}$, $G$, $P_{\rm sat}$, and the dark detector noise limit is determined by thermal fluctuation noise from phonon-mediated heat flow, NEP$_\mathrm{G}$, given by
\begin{linenomath*}
\begin{equation}
\mathrm{NEP}^2_\mathrm{G} = 4F_{\rm link}k_{\rm B}T^{2}_{\rm c}G,
\label{eq:nep_G}    
\end{equation}
\end{linenomath*}
where $k_{\rm B}$ is the Boltzmann constant and the factor $F_{\rm link}$ is classically approximated by~\cite{Boyle:59,Mauskopf_2018}
\begin{equation}
F_{\rm link} = \bigg[\frac{T_{\rm c}^{n+1} + T_{\rm b}^{n+1}}{2T_{\rm c}^{n+1}}\bigg].
\label{eq:nep2}    
\end{equation}

Array~2 was measured to have a somewhat higher $T_{\rm c}$ than the design target, while the saturation power $P_{\rm sat}$ for Array~1 was lower than the design target. The measured dark NEPs for Array~2 were consistent with the analytical estimate of NEP$_{G}$ from Eq.~\ref{eq:nep_G} and \ref{eq:nep2}. It may be noted that dark measurements of Array~1 prior to the final fabrication step of depositing the bismuth absorber layer yielded parameters closer to the design target. Since $\kappa$ and $n$ are related to the geometry of the legs, while $T_{\rm c}$ is the transition temperature for the superconducting film, no correlations were anticipated between $n$ or $\kappa$ and $T_{\rm c}$. We are investigating whether the lower than expected $P_{\rm sat}$ for Array~1 can be attributed to a slight misalignment in the mask used for depositing the bismuth absorber layer resulting in bismuth being deposited on the TES legs; see Fig.~\ref{fig:arr_imag}. 

\begin{figure*}
\begin{center}
\includegraphics[width=0.9\linewidth,keepaspectratio]{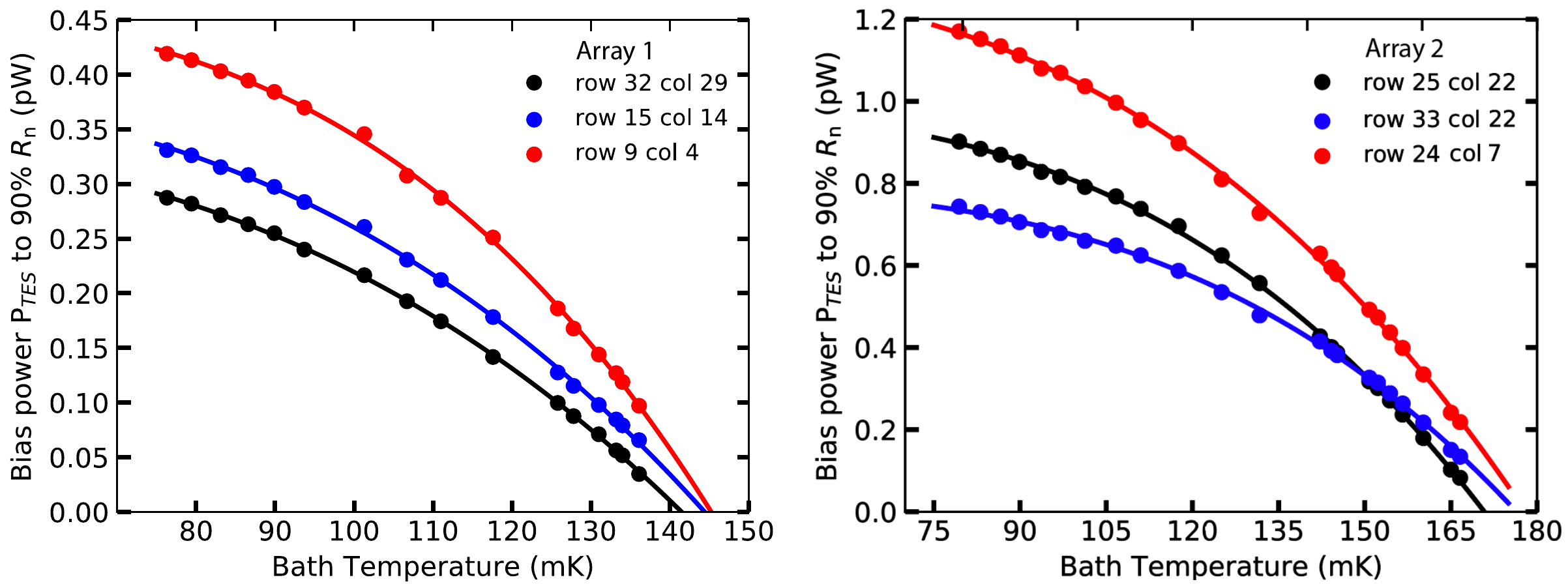}
\end{center}
\caption{Saturation power, $P_{\rm sat}$ as a function of bath temperature for three representative detectors each from Arrays 1 (left) and 2 (right) derived from calibrated IV curves acquired during pre-flight testing with the fully integrated receiver submerged in a LHe test dewar simulating conditions at float. The circles represent measured $P_{\rm sat}$ at various bath temperatures, and the lines represent the best-fit lines. The receiver window was covered with a sheet of Eccosorb maintained at ${\sim} 1.5$\,K using superfluid helium pumps~\cite{Kogut2021}. }
\label{fig:Arr_psat}
\end{figure*}

\begin{table}
\caption{Summary of detector array yield.}
\begin{center}
\begin{tabular}{lclclclcl}
\hline
Array && row/col failures  && random dead pixels  && overall yield \\ %$\tau (\mu$s) &&
\hline
1 && 14\% && 8\% && 78\% \\
2 && 17\% && 11\% && 72\% \\
\hline                 
\end{tabular}\\
\end{center}
\label{tab:yield}
\end{table}

Since Array~1 has low saturation power, a neutral density filter (NDF) is installed in front of both arrays to block a significant fraction ($\sim 70\%$) of the incident power in the observing band. The NDF is realized by depositing a thin layer of bismuth equivalent to a surface impedance of 220~$\Omega$/sq at a temperature of 4.2~K on a 0.5~mm thick $z$-cut quartz wafer, which provides a power transmission of $\sim$0.3. The total power loading on the detectors assuming an atmospheric contribution of 20~fW at a float altitude of 35~km is expected to be $\sim0.2$~pW in the PIPER 200~GHz band with a 30\% fractional bandwidth. Therefore, this NDF is a conservative but safe choice to avoid device saturation in the first science flight of PIPER. While it introduces a noise penalty, it mitigates risk by ensuring that the detectors will operate in transition in the unexpected case of higher instrumental (thermal) loading.  Once we have a measurement and verification of the total optical loading at float, the NDF could be relaxed or entirely removed. 

\subsection{Integrated Receiver Tests}
\label{subsec:int_test}

The detectors were tested after final integration with the PIPER receiver cryostat. Fig.~\ref{fig:det_int} shows a hybridized detector array packaged and ready for integration. The receiver integration and sub-Kelvin cooling are described in [\!\!\citenum{2019RScI...90i5104S}]. The integrated receiver was submerged in a custom dewar, partially filled with LHe and pumped out to simulate the ambient pressure at float. 

The receiver windows, which are located in the optical path after the analyzer grid when viewed from the sky, were capped with Eccosorb sheets cooled with superfluid helium pumps~\cite{Kogut2021} to remain isothermal to the pumped LHe bath. TES IV curves were acquired in this configuration and repeated at a few different Eccosorb temperatures obtained by controlling the superfluid helium pumps. With the CADR maintained at $\sim$~100~mK, we were able to keep the detectors in their transitions while looking at the Eccosorb sheet covering the window. Fig.~\ref{fig:Arr_psat} shows measured $P_{\rm sat}$ as a function of bath temperature along with best-fit lines for three representative detectors each from Arrays 1 and 2 derived from calibrated IV curves acquired during pre-flight testing.

\begin{figure}
\begin{center}
\includegraphics[width=1\linewidth,keepaspectratio]{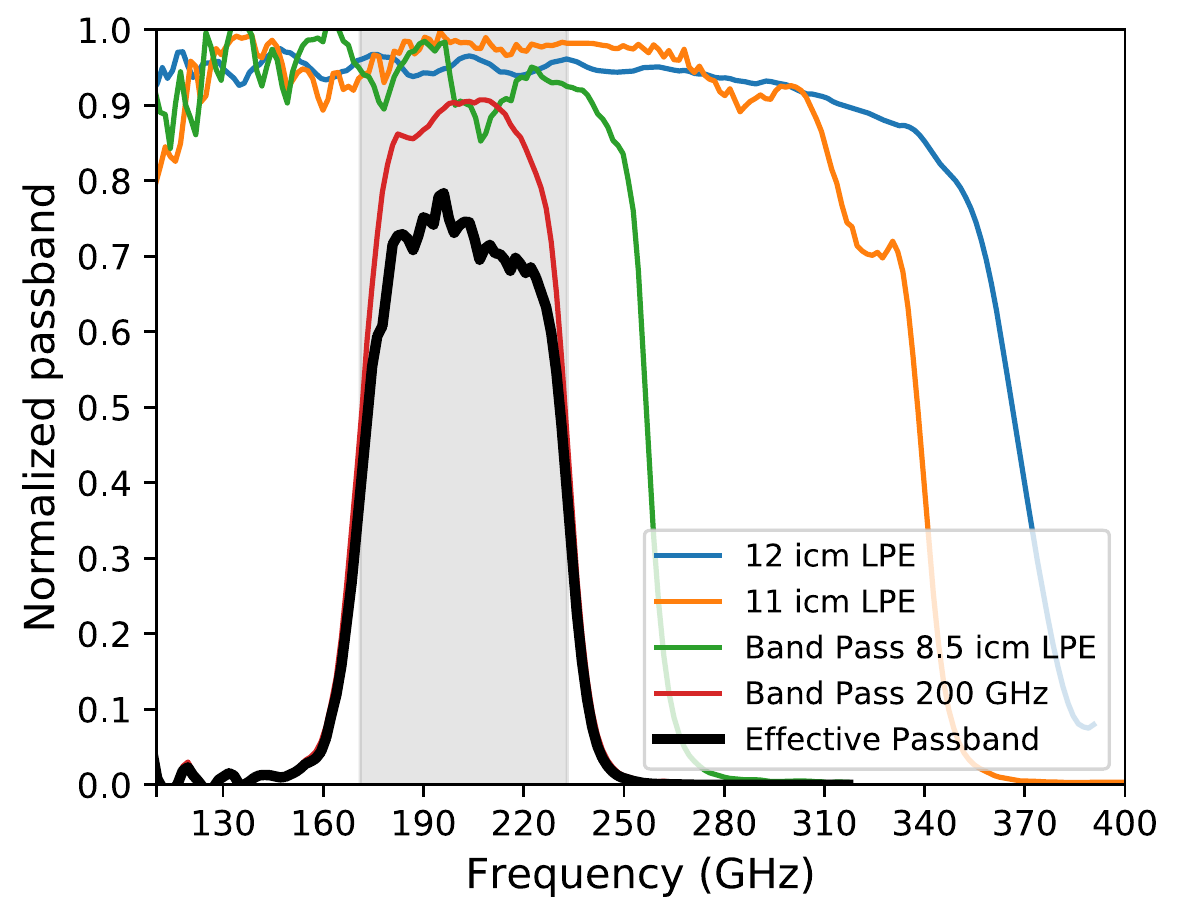}
\end{center}
\caption{The measured transmittance of the various filters and the effective filter-stack and window transmittance (in black) for the PIPER instrument’s 200 GHz channel. The grey band spans 171--233~GHz and represents the half-power bandwidth.}
\label{fig:200band}
\end{figure}

Assuming the cooled Eccosorb behaves like a blackbody load and fills the view of the detector, the expected power absorbed by a detector depends on the load temperature $T_{\rm L}$, its emissivity $\epsilon(\nu)$, and effective detector bandpass $f(\nu$) as
\begin{equation}
P_{\rm opt}(T_{\rm L})= \int{\epsilon(\nu)f(\nu)}A_{\rm e}(\nu)B(\nu,T_{\rm L}) d\Omega d\nu,
\label{eq:popt_full}    
\end{equation}
where B($\nu,T_{\rm L}$) is the Planck function, and A$_{\rm e}(\nu)$ is the effective area of the pixel which integrates over solid angle to $\lambda^{2}$ for each mode of radiation, i.e., the throughput (or etendue) of the instrument 
$A_{\rm e}\Omega = \lambda^{2}$ in the single-mode limit. While the multi-moded response of the detectors will need to be considered for the higher frequency PIPER channels, this single-mode limit is appropriate for the 200~GHz channel. Equation~\ref{eq:popt_full} then reduces to
\begin{equation}
P_{\rm opt}(T_{\rm L})= \int{\epsilon(\nu)f(\nu)} \frac{2h\nu}{e^\frac{h\nu}{kT_{\rm L}}-1} d\nu.
\label{eq:Popt_red}
\end{equation}

\begin{figure}
\begin{center}
\includegraphics[width=1\linewidth,keepaspectratio]{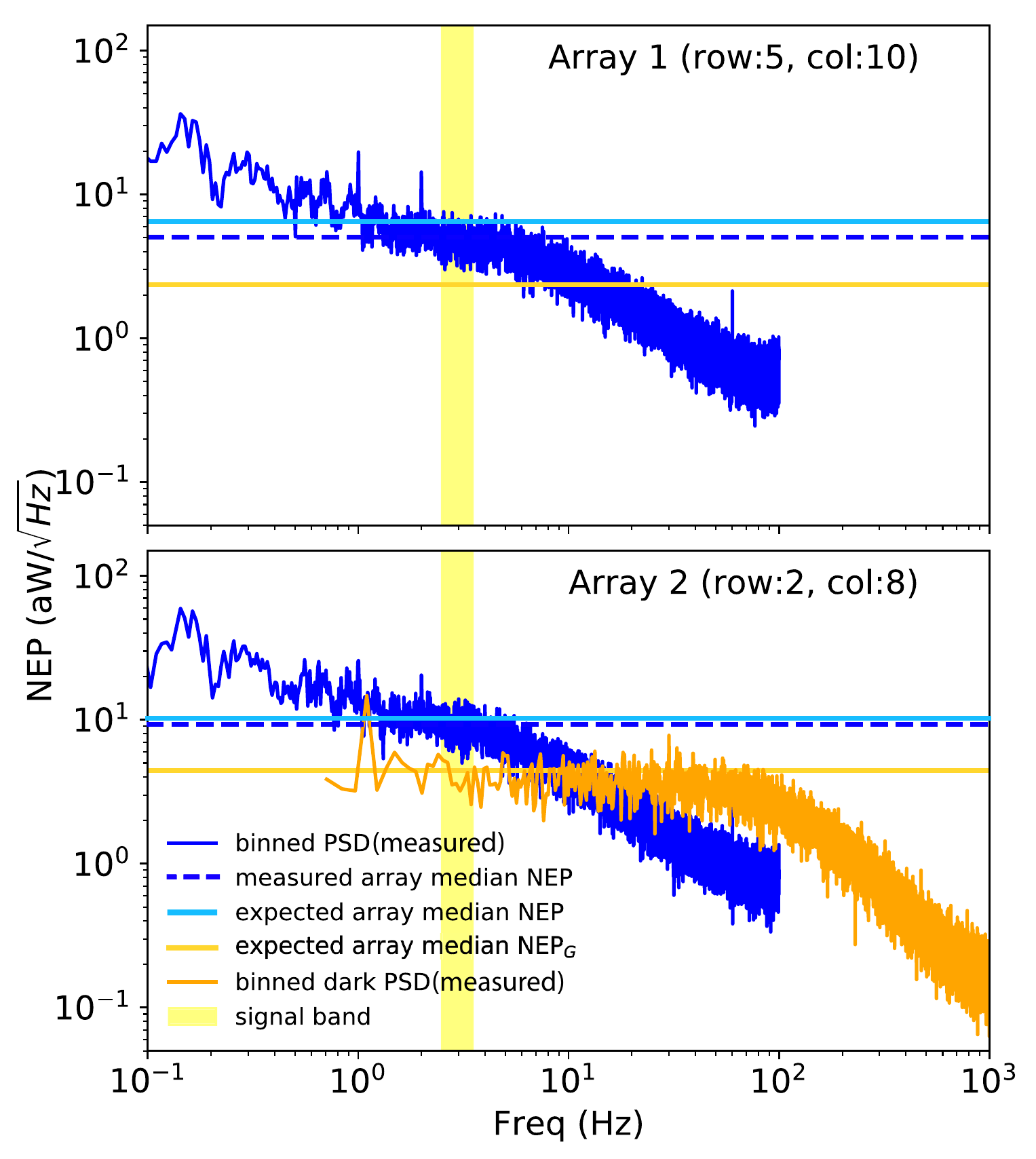}
\end{center}
\caption{Measured noise spectrum (blue) of a representative detector each from Arrays 1 and 2 biased in the transition in the integrated receiver test setup staring at an Eccosorb sheet covering the cryostat window and held at ${\sim} 2$\,K. The measured dark noise spectrum (orange, for Array~2 only) of the corresponding detector obtained from the dark test is also shown. The light blue solid lines show the expected array median NEP computed using Equation~\ref{eq:nep}, the blue dashed lines show the measured array median NEP. The orange solid lines show the expected array median NEP$_{G}$. The yellow highlighted band represents the signal band around the modulation frequency. }
\label{fig:nep_window}
\end{figure}

Assuming an emissivity~\cite{1985InfPh..25..561B} of the cooled Eccosorb $\epsilon(\nu)\sim 1$ independent of frequency over our range of interest, and $f(\nu)$ as the product of expected passband response (see Fig.~\ref{fig:200band} and a frequency-independent NDF transmission of 0.3, we can compute the expected power at the detector for a given Eccosorb temperature, $T_{L}$. For a given $P_{\rm opt}$, the noise-equivalent power (NEP) for an unpolarized detector can be modeled as a combination of the phonon noise and the photon noise
\begin{equation}
\mathrm{NEP}^2 = \mathrm{NEP}^2_\mathrm{dark} + 2h\nu_0 P_{\rm opt} + \frac{P^2_{\rm opt}}{\Delta\nu},
\label{eq:nep}    
\end{equation}
where $h$ is the Planck constant, and $\nu_0$ and $\Delta\nu$ are the detector center frequency and bandwidth, respectively. It should be noted here that the $P$ in the NEP above refers to absorbed optical power, which is distinct from power on the sky. 

The noise spectra obtained from 40~minute timestreams are shown for a representative detector from each array in Fig.~\ref{fig:nep_window} with a $\sim$~2~K Eccosorb covering the receiver window. Assuming the detectors were viewing a 2~K blackbody, the measured NEP was consistent with the total expected NEP. The bottom panel of Fig.~\ref{fig:nep_window} also shows the measured dark NEP for the same Array~2 pixel, which was consistent with the expected NEP$_{\rm G}$ given the measured thermal conductance, $G$. Table~\ref{tab:params} summarizes the measured array averaged thermal parameters and NEP.

\begin{table}
\caption{Mean and standard deviation of measured detector dark parameters, $T_{\rm c}$, $G$, and $P_{\rm sat}$. The NEPs are estimated from pre-flight integrated receiver tests, and phonon noise estimates (NEP$_{\rm G}$) from dark measurements of the conductance $G$ are noted in parenthesis.}
\begin{center}
\begin{tabular}{lclclclclcl}
\hline
Array && $T_\mathrm{c}$(mK) && $G$ (pW/K) && $P_\mathrm{\rm sat}$(pW) && NEP (aW/$\sqrt{Hz}$)\\ %$\tau (\mu$s) &&
\hline
1 && 147.9$\pm$5.7 && 8.3$\pm$0.8 && 0.26$\pm$0.04 && 5.0$\pm$1.2 (2.4)\\
2 && 179.9$\pm$3.7 && 18.9$\pm$3.3 && 0.8$\pm$0.14 && 9.4$\pm$2.1 (4.5)\\
\hline                 
\end{tabular}\\
\end{center}
\label{tab:params}
\end{table}

\section{Summary}
\label{sec:conclusion}

We present measurements of dark detector parameters and noise measurements from integrated receiver tests of 32$\times$40 pixel PIPER BUG TES detector arrays hybridized with two-dimensional SQUID-based chips implementing time division multiplexing. The detectors demonstrate low saturation powers ($<1$~pW) and low noise (few aW/$\sqrt{\rm Hz}$), meeting requirements for the PIPER balloon-borne instrument. For one of the two arrays, during the final absorber deposition step to maximize detector coupling efficiency, the mask was slightly misaligned, resulting in the Bi absorber overlapping with the TES legs, lowering the detector saturation powers below the design target between pre- and post- absorber deposition. The achieved saturation powers for this array are marginal but acceptable. The detectors are currently integrated with the PIPER receiver and flight-ready. 
%This detector technology could also be relevant for future long wavelength end of far-IR or CMB observatories.

\begin{acknowledgments}
S.D. is supported by an appointment to the NASA Postdoctoral Program at the NASA Goddard Space Flight Center, administered by Oak Ridge Associated Universities under contract with NASA. PIPER is supported by grant 13-APRA13-0093. The GSFC Internal Research and Development (IRAD) program provided support for the development of microwave instrumentation and metrology. R. D.'s research was supported by an appointment to the NASA Postdoctoral Program at the NASA Goddard Space Flight Center (GSFC), administered by Universities Space Research Association under contract with NASA. The authors would like to thank Shawn W. Henderson and Jason R. Stevens for their help with implementing the SQUID multiplexer chip screening procedure.

\end{acknowledgments}

\cite{*}

\bibliography{references}

\end{document}